\documentclass[12pt, letterpaper]{article}
\usepackage{geometry}
\usepackage{amsmath}
\usepackage{graphicx}
\usepackage{float}
\usepackage{subcaption}
\usepackage{amssymb}
\newgeometry{vmargin={18mm}, hmargin={18mm,18mm}}
\begin{document}
\title{Adiabatic evolution of solitons embedded on \\ lipid membranes}
\author{O. Pav\'on-Torres$^{a,} \footnote{Corresponding author: omar.pavon@cinvestav.mx}$, M. A. Agüero-Granados$^{b}$, R. Valencia-Torres$^{a}$}
\date{%
\small{$^{a}$ Physics Department, Cinvestav, POB 14-740, 07000 Mexico City, Mexico\\ %
$^{b}$ Facultad de Ciencias, Universidad Aut\'onoma del Estado de M\'exico, \\ Instituto literario 100, Toluca 5000, M\'exico.\\
}}
\maketitle

\begin{abstract}
The Heimburg-Jackson model, or thermodynamic soliton theory of nervous impulses, has a well-established record as an alternative model for studying the dynamics of nerve impulses and lipid bilayers. Within this framework, nerve impulses can be represented as nonlinear excitations of low amplitude depicted by the damped nonlinear Schrödinger equation and their adiabatic evolution can be analyzed using direct perturbative methods. Based on the foregoing, we carry out the current study using the quasi-stationary approach to obtain the adiabatic evolution of solitons embedded in lipid bilayers under the influence of a viscous elastic fluid. This analysis encompasses liquid-to-gel transition of the lipid bilayers, for whose dark and bright solitons arise, respectively. 
\end{abstract}

\section{Introduction}
 
Neurons are complex structures responsible for generating and propagating nerve impulses. In biological terms, neurons are cellular structures enveloped by lipid membranes. These membranes are the sites where most processes involved in neuronal preservation and functioning are triggered. Consequently, many neuropathological phenomena originate from this structure due to membrane dysfunction \cite{0, 01, 02}. Operationally, the neuronal membrane resembles a lipid bilayer supported by a hydrophobic surface submerged in water \cite{r0000}. Lipid bilayers posses intriguing thermodynamic properties and their models are essential in our current understanding of bio-membranes. For instance, lipid bilayers composed of a single type of lipid undergo a phase transition where they melt at a specific temperature. Above this temperature, they exist in a two-dimensional disordered liquid phase, while below it, they adopt an ordered gel phase \cite{r001}. This property is crucial for comprehending the behavior and evolution of nerve impulses, as well as sensory stimuli and numerous neurophysiological processes.

Historically, the nerve impulses were first conceived as a pure electrical phenomenon, in which the surface membrane is modelled as an electric circuit. Within this framework, the cell membrane assumed the role of a capacitor, while ion channels functioned akin to resistors. As the ion channels open and close in time, ionic currents flow across the membrane generating an alteration of the transmembrane voltage or action potential (AP), which spreads over long distances. The HH model \cite{r00, r01} and the FitzHugh-Nagumo (FHN) model \cite{nag1, nag2} have been fundamental in modelling these electrical phenomena, offering distinct yet complementary perspectives on neural dynamics.  Nevertheless, despite being one of the most studied models, the HH model do not offer a whole description of the nerve impulses. The HH model does not incorporate the non-electrical aspects of nerve impulses such as mechanical, thermal and optical aspects, for which there have been a vast amount of empirical evidence throughout the years \cite{cris1, cris2, cris3, cris4}. To incorporate the mentioned features of nerve impulses, a lot of contemporary studies have been developed implementing interdisciplinary approaches \cite{heim1, heim2, heim3, heim4, maxom0, r2, r6, r000, r0, ji0, ji1, ji2}. 

Among the most famous models, the Heimburg Jackson model or the so-called thermodynamic soliton has underscored.  Such a model assumes that the nervous impulse is an adiabatic density pulse during which the membrane changes area, thickness, and temperature. The soliton theory is of macroscopic thermodynamic nature ensuring its consistent with the laws of thermodynamics \cite{heim1, heim2, heim3, heim4}. In this model the propagation of a solitary wave can be achieved by bringing a tiny section of the fluid membrane into gel state phase. Indeed, near the phase transition a local change of temperature, a sudden change in pH, a local increase in calcium concentration or a electrostatic potential, could trigger this pulse.  One of the many appeals of such theory is the possibility of explaining the orthodromic and antidromic impulse annihilation and their evolution in the axoplasmic fluid \cite{cinderella1}. 
Despite all the alternatives and possibilities that this models grants, it presents some apparent experimental inconsistencies regarding the propagation of nerve impulses \cite{o2, maxom1}. For instance, the phase transition close to a physiological temperature in the axon membrane has not been experimentally reported. Nevertheless, such a transition has been observed in unilamellar vesicles, bovine lung surfactant and two bacterial membranes\cite{oh1}. In addition, and respect to the orthodromic-antidromic annihilation upon collision, Follman et al.\cite{yoyoi3} used more intricate designs attempted to replicate the findings of Gonzalez-Perez et al. \cite{yoyoi1, yoyoi2}, but observed no soliton like behaviour, such as reflection or crossing of the action potentials.
Even these last observations seem to be discouraging, recently experiments with lipid bilayers in which membranes were locally heated within a tiny region and they observe that far away from such a region,the arrival of a mechanical perturbation were optically measured, presumably triggered by the thermal input \cite{o1, r5, oo4, oo5, 00}. Supporting the idea of a thermodynamic soliton theory and its role in real neurons. In the framework of the weak formulation of the Heimburg-Jackson model, Contreras et al. \cite{r9, r17} obtained analytically two forms of solutions for the density equation. The gel state of the nerve, the soliton model admits bright solitons, whereas in the liquid states they admit dark solitons. Similar soliton-like profiles can be obtained directly starting from the diffusion equation of the Heimburg-Jackson model and reducing it to a damped nonlinear Schrödinger equation \cite{t1}. This approach allows to work directly with the nerve impulses profiles that appears in gel and liquid state of nerve to study their adiabatic evolution as they propagate along the nerve fibre and the effect of the intracellular fluid or axoplasmic fluid on these structures by means of perturbative methods. Thus, the following work is going to be divided in the following sections. Section 2 is devoted to present an overview of the Heimburg-Jackson model of nervous system as well as the reduction procedure that yields to the damped nonlinear Schrödinger equation. In section 3, we employ the quasi-stationary approach to study the adiabatic evolution of bright and dark solitons in the gel and liquid state of the nerve due to the effect of the elastic viscous medium (axoplasmic fluid). Finally, the conclusion section is devoted to discuss our main results and provide a wide perspective analysis. 

\section{Thermodynamic soliton theory of the nervous impulse}

A myelinated nerve fibre is composed of an insulating protein tube called the axolemma, which has a lipid bilayer structure. Each layer consists of amphiphilic phospholipids with hydrophilic heads and hydrophobic tails. The interior of the nerve fiber is filled with a conductive electrolyte known as axoplasm, or intracellular fluid. Axoplasm contains cytoskeletal filaments and a specific concentration of ions, making it a viscous medium from a mechanical perspective (for more details on the biological aspects of nerve fibres, refer to \cite{coo}).

\begin{figure}[ht]
\centering
\includegraphics[width=7cm]{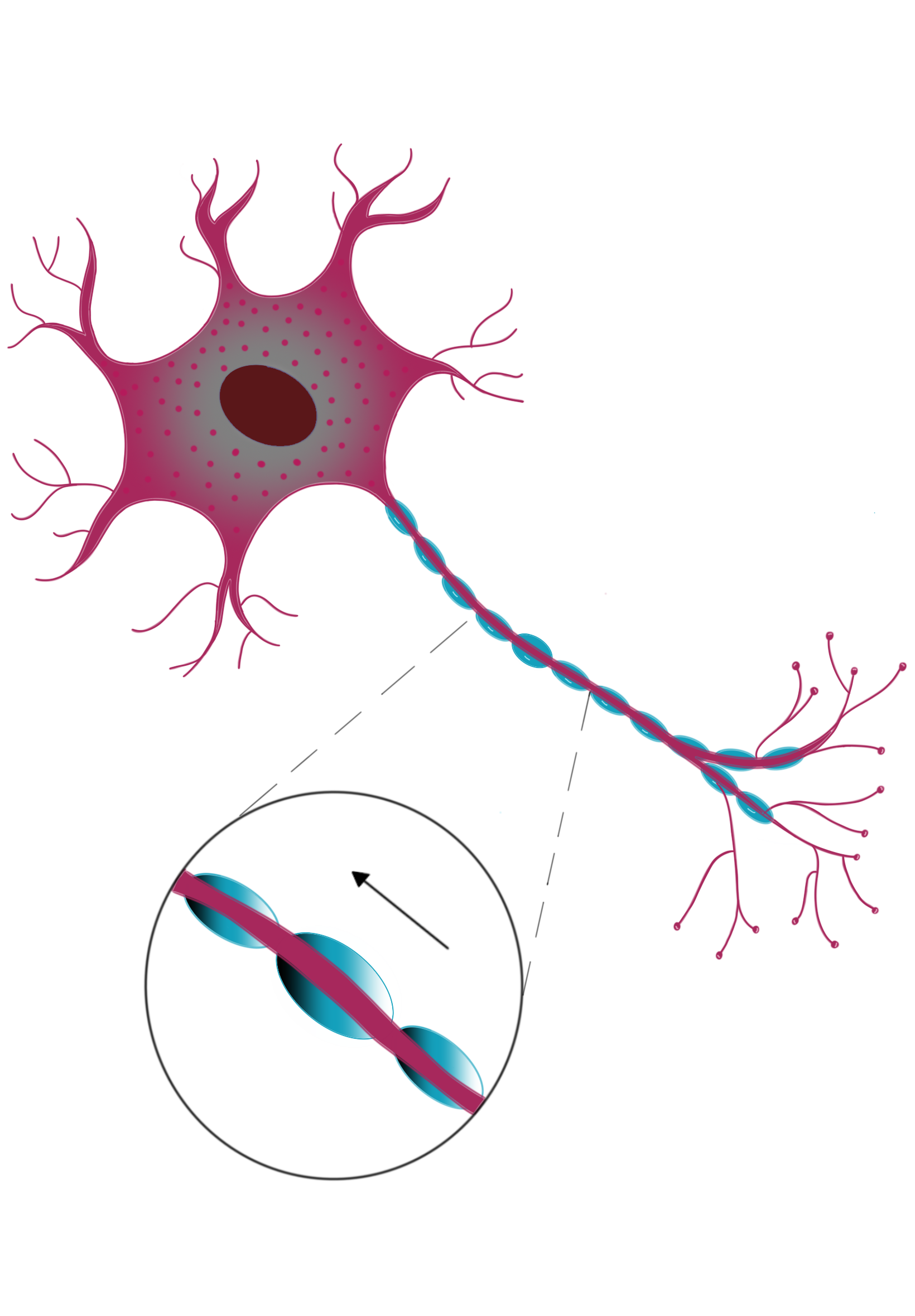}
\caption{Representation of mechanical distortion of nerve.} \label{ig1}
\end{figure}

The lipid bilayer structure, or biomembrane, contains voltage-gated ion channels responsible for regulating ion flow between the intracellular fluid and the extracellular environment. The concentrations of ions in these fluids play a critical role in generating nerve signals. As mentioned earlier, this transmembrane voltage is known as the action potential (AP) and is initiated at the axon hillock within neurons by an electrical impulse exceeding a certain threshold amplitude. However, action potentials can also be triggered at remote sites from the axon hillock. The propagation of the action potential follows two primary pathways: orthodromic, where impulses travel from the soma towards synaptic terminals, and antidromic, where propagation occurs in the opposite direction. The AP travels through the axoplasm, supported by ion currents across the biomembrane. Experimentally, action potential microscopy allows for the visualization of AP propagation along nerve fibers. This method employs fluorescent molecular probes sensitive to local electric fields. These probes are introduced into the axoplasm, and the nerve fiber is electrically stimulated in a standard manner. At the peak of the action potential, a bright fluorescence spot appears, which can be monitored in real time using optical techniques and a digital camera \cite{gogogoi}. As the action potential propagates along the nerve fiber, it induces a phase transition and mechanical distortions in the insulating protein tube of the nerve \cite{cris2, cris3}. Thus, experimental evidence shows that the electrical stimulation increases the diameter of nerve fibres (as show in Fig. \ref{ig1}) as by significant temperature changes. 

\subsection{The Heimburg-Jackson model}

To incorporate mechanical and thermodynamic effects in nerve fibres, we follow the basic ideas of Heimburg and Jackson. Thus, we start from the wave equation 
\begin{equation}
\dfrac{\partial^{2}U}{\partial t^{2}}=\dfrac{\partial}{\partial x}\left(c^ {2}\dfrac{\partial U}{\partial x}\right)\label{panda00},
\end{equation}
where $U=\rho^ {A}-\rho_{0}^ {A}$ is the density change between the gel-state density $(\rho^{A})$ and the fluid-state density $(\rho^{A}_{0})$. 
The first assumption of the Heimburg-Jackson model establishes a fundamental connection between the velocity $c$ and the compressibility inherent in the lipid composition of the circular biomembrane. It is assumed that
\begin{equation}
c^{2}=c_{0}^ {2}+pU+qU^ {2},\label{panda0}
\end{equation}
where $c_{0}$ is the speed of sound during the phase transition, $p$ and $q$ are nonlinear elastic properties of the membrane determined from the experiments as follows
\begin{equation}
c_{0}^ {2}=\dfrac{1}{K_{s}^ {A}\rho_{0}^ {A}}; \quad p= -\dfrac{1}{\left(K_{s}^ {A}\rho_{0}^ {A}\right)^ {2}}; \quad q= \dfrac{1}{\left(K_{s}^ {A}\rho_{0}^ {A}\right)^ {3}},
\end{equation}
being $K_{s}^ {A}$ the lateral compressibility of the axon. Replacing Eq. (\ref{panda0}) in Eq. (\ref{panda00}) the Heimburg-Jackson model turns out to be
\begin{equation}
\dfrac{\partial^{2}U}{\partial t^{2}}=\dfrac{\partial}{\partial x}\left[\left(c_{0}^ {2}+pU+qU^ {2}\right)\dfrac{\partial U}{\partial x}\right]-h\dfrac{\partial ^ {4}U}{\partial x^ {4}},  \label{panda}
\end{equation}
with the added higher order $-h\partial^{4} U/\partial x^{4}$ term being responsible for dispersion. The study of solutions to the diffusion equation (\ref{panda}) involving Jacobi elliptic functions and soliton-like solutions represents an intriguing and ongoing research area \cite{r7, multi, multi2, r8, r4}. From these solutions the multiwaves, breathers and lump solitons are worth of noticing. We are particularly aimed on studying the axoplasmatic effect, or in mechanical terms, the impact of friction on soliton-like solutions. To account for this effect, we introduce a mixed derivative term into Equation (\ref{panda}), resulting in:
\begin{equation}
\dfrac{\partial^{2}U}{\partial t^{2}}=\dfrac{\partial}{\partial x}\left[\left(c_{0}^ {2}+pU+qU^ {2}\right)\dfrac{\partial U}{\partial x}\right]+\alpha\dfrac{\partial ^ {2}}{\partial x^ {2}}\left(\dfrac{\partial U}{\partial t}\right)-h\dfrac{\partial ^ {4}U}{\partial x^ {4}}, \label{panda1}
\end{equation}
the incorporation of the additional term is possible due to the similarity between the nerve fibre and the rods. A similar term is responsible for inertial effects in rods and it can be related to inertia of the lipids in biomembranes \cite{r0}. 

\subsection{The damped nonlinear Schrödinger equation}
Expressing Eq. (\ref{panda1}) in terms of dimensionless variables we have

\begin{equation}
\dfrac{\partial^{2}u}{\partial \tau^{2}}=\dfrac{\partial}{\partial \tilde{z}}\left[\left(1+\tilde{p}U+\tilde{q}U^ {2}\right)\dfrac{\partial u}{\partial \tilde{z}}\right]+\mu\dfrac{\partial ^ {2}}{\partial \tilde{z}^ {2}}\left(\dfrac{\partial u}{\partial \tau}\right)-\dfrac{\partial ^ {4}U}{\partial \tilde{z}^ {4}}, \label{panda2}
\end{equation}
where
\begin{equation}
u=\dfrac{U}{\rho_{0}^{A}}, \quad \tilde{z}=\dfrac{c_{0}x}{\sqrt{h}}, \quad \tau=\dfrac{c_{0}^{2}t}{\sqrt{h}}
\end{equation}
with the new parameters
$$\mu=\dfrac{\alpha}{\sqrt{h}}, \quad \tilde{p}=\dfrac{\rho_{0}^{A}}{c_{0}^{2}}p \quad \tilde{q}=\dfrac{\left(\rho_{0}^{A}\right)^ {2}}{c_{0}^{2}}q.$$
Orthodromic as well as antidromic impulses can be modelled as low amplitude nonlinear excitations in the weakly dissipative soliton model. Thus, we reduce the Heimburg-Jackson diffusion equation to a damped nonlinear Schrödinger equation (see \cite{t1} and references therein). We introduce a small parameter $\epsilon\ll 1$, and substitute $\tilde{p} \to \epsilon^{2}\tilde{p}$, $\tilde{q} \to \epsilon^{2}\tilde{q}$, $\mu \to \epsilon \mu$ in Eq. (\ref{panda2}).  Followed by a change of variables $y=\epsilon (\tilde{z}-\tau)$ and $s=\epsilon^{3}\tau$ we obtain the Burgers-Korteweg-de Vries (BKdV) equation
\begin{equation}
\dfrac{\partial u}{\partial s}+\dfrac{1}{2}(a\tilde{p}u+a^{2}\tilde{q}u^{2})u_{y}-\dfrac{1}{2}\dfrac{\partial^ {3}u}{\partial y^{3}}=a^{2}\dfrac{\mu}{2}\dfrac{\partial^ {2}u}{\partial y^{2}}. \label{ar1}
\end{equation}
We could start our main analysis of solitons embedded in lipid membranes from the BKdV equation, performing the direct perturbation analysis in a similar fashion as it will be done in the following, nevertheless the damped nonlinear Schrödinger equation as a prototypical model in many physical situations turns a more used model. Thus, the mathematical approaches introduced in the present study can be extrapolated to wider nonlinear phenomena. A direct approach in the study of soliton perturbations in the BKdV equation can be a complete discussion in \cite{homer1}.
We use a multiple-scale expansion method \cite{mirage1, r10}. The method involves introducing two time and position scales, i.e., the fast time and position scales for oscillations and the slow time and spatial scales for the envelope amplitude. Thus, we introduce new time scale $s_{i}=a^ {i}s$ and space scale $y_{i}=a^ {i}y$, and assume a solution in terms of the Poincar\'e type expansion
\begin{equation}
u(y, s)=\sum_{i=0}^ {\infty}a^{i}u(s_{0}, s_{1}, s_{2}, y_{0}, y_{1}),
\end{equation}
in which $s_{i}$ and $y_{i}$ is treated as an independent variable. This leads to a perturbation series of operators for all independent variables 
\begin{equation}
\dfrac{\partial}{\partial s}=\dfrac{\partial}{\partial s_{0}}+a\dfrac{\partial}{\partial s_{1}}+a^ {2}\dfrac{\partial}{\partial s_{2}}+\dots \label{wom1}
\end{equation}
The solution of the original problem will be obtained for the new sets of variables $s_{i}$ and $y_{i}$ with the explicit form given by
\begin{equation}
s_{0}=s, \quad s_{1}=as, \quad s_{2}=a^ {2}s. \label{ladworm}.
\end{equation}
Replacing the above operator (\ref{wom1}) and their counterpart for $y$ into the different terms of the BKdV equation and group the terms of the same order of $a$ to obtain a system of equations. Each of these equations will correspond to each approximation having harmonics of a specific order. Grouping the operators:

\begin{equation}
\dfrac{\partial^ {2}}{\partial y^{2}}=\dfrac{\partial^{2}}{\partial y_{0}^{2}}+2a\dfrac{\partial^ {2}}{\partial y_{0}\partial y_{1}}+a^{2}\dfrac{\partial^ {2}}{\partial y_{1}^{2}},\label{wom3}
\end{equation}
and writing $u$ as a perturbative series, and consider only terms to the first order of $a$, i.e., 
\begin{equation}
u=Ae^{i\theta}+A^{*}e^{-i\theta}+a(C+Be^{2i\theta}+B^{*}e^{-2i\theta}), \label{wom4}
\end{equation}
where $\theta=(ky_{0}-\omega s_{0})$ and the amplitudes $A$, $B$ and $C$ correspond to $(s_{1}, y_{1}, y_{2})$. Substituting Eqs. (\ref{wom1})-(\ref{wom4}) in Eq. (\ref{panda2}), and looking for relations between terms of same orders in $a$ with terms in $e^{\pm i \theta}$, $e^{\pm 2i \theta}$ without an exponential dependence set to zero. To the order $a^{0}$ the annihilation of terms in $e^ {\pm i \theta}$ gives the dispersion relation of linear waves, i.e., 
\begin{equation}
\omega=\dfrac{k^{3}}{2}.\label{frec}
\end{equation}
To the order $a^{1}$ we have
\begin{equation}
\dfrac{\partial A}{\partial s_{1}}+v_{g}\dfrac{\partial A}{\partial y_{1}}=0 \label{lw1},
\end{equation}
where $v_{g}$ is the group velocity defined as
\begin{equation}
v_{g}=\dfrac{\partial \omega}{\partial k}=\dfrac{3k^{2}}{2}.\label{vg}
\end{equation}
Setting terms in $e^{\pm 2i \theta}$ to zero yield to 
\begin{equation}
B=-\dfrac{pk}{6k^{3}}A^{2}. \label{wwmo0}
\end{equation}
To the second order, the terms with no exponential dependence are
\begin{equation}
\dfrac{\partial C}{\partial s_{1}}-\dfrac{p}{2}\dfrac{\partial |A|^ {2}}{\partial y_{1}}=0.
\end{equation}
Now, considering the transformations $y=(y_{1}-v_{g}s_{1})$ and $\tau=s_{1}$, and comparing the result with Eq. (\ref{lw1})  we obtain
\begin{equation}
C=\dfrac{p}{3k^ {2}}|A|^{2} \label{wwom1}
\end{equation}
To the second order perturbation,  the terms of order $e^{\pm i \theta}$ finally give the damped nonlinear Schrödinger equation we were looking for
\begin{equation}
i\dfrac{\partial A}{\partial s_{2}}+P\dfrac{\partial^{2} A}{\partial y_{1}^{2}}+Q|A|^{2}A+iRA=0 \label{schr1},
\end{equation}
which describes the evolution of the envelope amplitude of the density pulse $u$ in the one-dimensional cylindrical nerve axon. Here the nonlinearity or (self-trapping), the damping and the dispersive coefficients $Q$, $R$ and $P$, respectively, are real and defined in terms of membrane parameters as
\begin{equation}
Q=\left(\dfrac{\rho_{0}^{A}}{2c_{0}}\right)\left(\dfrac{p^{2}}{3c_{0}^{2}k^{2}}-k{q}\right), \quad P=\dfrac{3k}{2}, \quad R=\dfrac{\alpha k^ {2}}{2\sqrt{h}}\label{agon0}.
\end{equation}
The inclusion of the imaginary term in the nonlinear Schrödinger equation is keynote in depicting the damping of amplitude, illustrating the irreversible nature of time evolution. This damping term is often attributed to the elastic viscous properties of the axoplasm. Bright and dark solitons, arise from the phase transitions of the lipid membrane \cite{r9, r17} for the idealized case when there is no surrounding medium. In general, $k$ in Eq. $(\ref{agon0})$ can assume either positive or negative values to obtain both soliton profiles. Under the present approach, the study of the pressure wave inside the cell membrane and its interaction with the axoplasmic fluid can be done straightforward. The NLSE appears in a many branches of mathematical physics as a prototypical model; consequently, its solutions have been exhaustively studied. The solutions for the undamped equation (case when $\gamma=0$ in Eq. (\ref{schr2})) have been exhaustively studied in literature. Solutions of particular interest are: bright and dark solitons, rogue waves, kink, antikink, lump solitons, among others \cite{jojoy0, jojoy1, jojoy2, jojoy3, jojoy4, jojoy5, homer0}.

\section{Adiabatic evolution of solitons in phase transitions of lipid bilayer}
\subsection{Adiabatic evolution of bright soliton in gel state of lipid bilayer}
For the sake of simplicity and to obtain the effect of the axoplasmatic fluid on solitons on the gel and liquid state of lipid bilayers, we work with normalized constants in Eq. (\ref{schr1}). By substituting $R$ for $\gamma$, $s_{2}$ for $\tilde{t}$ and $y_{1}$ for $z$, we have

\begin{equation}
i\dfrac{\partial A}{\partial  \tilde{t}}\pm \dfrac{1}{2}\dfrac{\partial^{2} A}{\partial z^{2}}+|A|^{2}A+i\gamma A=0 \label{schr2},
\end{equation}
Even the whole branch of solutions can be obtained from NLSE by a careful choice of the initial conditions, we are interested in the bright and dark soliton solutions due to the fact that they characterize the nerve impulse in the gel and liquid state and its evolution along the nerve channel. 

The one envelope soliton solution for the normalized nonlinear Schrödinger equation (\ref{schr2}) with $\gamma=0$, corresponding to a nonlinear excitation of low amplitude within nerve impulse in gel state of nerve, is given by  
\begin{equation}
A(z, t)=\rho \text{sech} \rho {(\varphi-\varphi_{0})}\exp[iv(\varphi-\varphi_{0})+i(\sigma-\sigma_{0})],\label{sol1}
\end{equation}
where
\begin{equation}
\dfrac{\partial \varphi}{\partial \tilde{t}}=-v, \quad \dfrac{\partial \varphi}{\partial z}=1, \quad \dfrac{\partial \sigma}{\partial \tilde{t}}=\dfrac{\rho^ {2}}{2}+\dfrac{v^{2}}{2} \quad \text{and} \quad \dfrac{\partial \sigma}{\partial z}=0. \label{f1}
\end{equation}

The bright soliton solution Eq.(\ref{sol1}) is fully characterized when the four parameters (\ref{f1}) are given. In this representation we have that $\rho$ represents the amplitude of the soliton as well as the pulse width of the wave, $v$ represents its speed which stands for a deviation from the group velocity as well as the frequency shift, $\varphi_{0}$ and $\sigma_{0}$ represent phase constants.   Following the standard procedure of the quasi-stationary approach, whose implementation has been successful employed in the description of another physical systems \cite{o5}. We introduce a slow time variable $T=\gamma \tilde{t}$, and treat the quantities $\rho$, $\sigma$, $\varphi_{0}$ and $\sigma_{0}$ as functions of this time scale. Hence, the envelope soliton solution (\ref{sol1}) is written as

\begin{equation}
A=\hat{A}(\varphi, T; \gamma)\exp[iv (\varphi-\varphi_{0})+i(\sigma-\sigma_{0})].
\end{equation} 

Under the assumption of quasistationarity, the perturbed nonlinear Schrödinger equation turns out to be 

\begin{equation}
-\dfrac{\rho^{2}}{2}\hat{A}+\dfrac{1}{2}\hat{A}_{\varphi \varphi}+|\hat{A}|^{2}\hat{A}=\gamma F({\hat{A}}) \label{max1}
\end{equation}
where 
\begin{equation}
F(\hat{A})=[(\varphi-\varphi_{0})v_{T}-v\varphi_{0T}-\sigma_{0T}]\hat{A}-i[\hat{A}_{T}+\hat{A}].\label{max2}
\end{equation}
Again, assuming a Poincar\'e-type asymptotic expansion $\hat{A}(\varphi, T; \gamma)=\sum_{n=0}^{\infty}\gamma^{n}\hat{A}_{n}(\varphi, T)$ and further restrict ourselves to the calculations of first order perturbations, i.e., $\hat{A}(\varphi, T; \gamma)=\hat{A}_{0}(\theta, T)+\gamma \hat{A}_{1}(\varphi, T)$, where $\hat{A}_{0}=\rho \text{sech}[\rho(\varphi-\varphi_{0})]$. Assuming $\hat{A}_{1}=\phi_{1}+i\psi_{1}$, where $\phi_{1}$ and $\psi_{1}$ are real and on substitution the above in Eqs. (\ref{max1}) and (\ref{max2}), we obtain 

\begin{subequations}
\begin{equation}
L_{1}\phi_{1}=-\dfrac{\rho^ {2}}{2}\phi_{1}+\dfrac{1}{2}\phi_{1\varphi\varphi}+3|\hat{A}_{0}|^{2}\phi_{1}=\text{Re}F(\hat{A}_{0})\label{max3}
\end{equation}
\begin{equation}
L_{2}\psi_{1}=-\dfrac{\rho^{2}}{2}\psi_{1}+\dfrac{1}{2}\psi_{1\varphi\varphi}+|\hat{A}_{0}|^{2}\psi_{1}=\text{Im}F(\hat{A}_{0})\label{max4}
\end{equation}
\end{subequations}
with
\begin{subequations}
\begin{equation}
\text{Re}F(\hat{A}_{0})=[(\varphi-\varphi_{0})v_{T}-v \varphi_{0T}-\sigma_{0T}]\hat{A}_{0},\label{max5}
\end{equation}
\begin{equation}
\text{Im}F(\hat{A}_{0})=-[\hat{A}_{0T}+\hat{A}_{0}].\label{max6}
\end{equation}
\end{subequations}

In Eqs. (\ref{max3}) and (\ref{max4}) we can notice that $L_{1}$ and $L_{2}$ are self-adjoint operators. Thus, the solutions of the homogeneous parts of Eqs. (\ref{max3}) and (\ref{max4}) are $\hat{A}_{0\varphi}$ and $A_{0}$, respectively, and hence we have the following secularity conditions:

\begin{subequations}
\begin{equation}
\int_{-\infty}^{\infty}\hat{A}_{0\varphi}\text{Re}F(\hat{A}_{0})d\varphi=0,
\end{equation}
\begin{equation}
\int_{-\infty}^{\infty}\hat{A}_{0}\text{Im}F(\hat{A}_{0})d\varphi=0.
\end{equation}
\end{subequations}
Integrating and substituting the values of $\hat{A}_{0\varphi}$, $\hat{A}_{0}$, $\text{Re}F(\hat{A}_{0})$ and in $\text{Im}F(\hat{A}_{0})$, we obtain $v_{T}=0$ and $\rho_{T}=-2\rho$, which can be written after integrating once as 
\begin{equation}
v=v_{0}, \qquad \rho=\rho_{0}e^{-2\gamma \tilde{t}}, \label{max7}
\end{equation}
where $v_{0}$ and $\rho_{0}$ are the initial velocity and amplitude of the soliton. The first equation (\ref{max7}) says that when the surrounding axoplasmic fluid interacts with the system, the velocity of the soliton is unaffected by it. Nevertheless, from the $\rho$ equation of (\ref{max7}), one understands that the viscous effect of the surrounding medium decreases the amplitude of the soliton excitations. The latters results in a exponential damping of the soliton excitations due to the surrounding axoplasmic fluid and hence it will travel only for a limited distance. The $\rho$ equation of Eq. (\ref{max7}) tells that the amplitude of the soliton reduces to $1/e$ times the initial amplitude $\rho_{0}$ after a duration of $T=1/2\gamma$, from when the perturbation due to the axoplasm effect is switched on. 
Now, we construct the perturbed soliton solutions by solving Eqs. (\ref{max3}) and (\ref{max4}). Thus, we have
\begin{eqnarray}
\phi_{1} &=& C_{1}\phi_{11}+C_{2}\phi_{12}-\phi_{11}\int_{-\infty}^{\infty}\phi_{12}\text{Re}F(A_{0})d\varphi\\ \nonumber
&+& \phi_{12}\int_{-\infty}^{\infty}\phi_{11}\text{Re}F(A_{0})d\varphi,\label{max8}
\end{eqnarray}
with 
\begin{subequations}
\begin{equation}
\phi_{11}=\text{sech}\rho (\varphi-\varphi_{0})\text{tanh} \rho (\varphi-\varphi_{0}), \label{max9}
\end{equation}
\begin{eqnarray}
\phi_{12} &=& \dfrac{1}{\rho}\bigg[\dfrac{3}{2}\rho(\varphi-\varphi_{0})\text{sech} \rho(\varphi-\varphi_{0})\text{tahn} \rho (\varphi-\varphi_{0}) \\ \nonumber
&+&\dfrac{1}{2}\text{tahn} \rho (\varphi-\varphi_{0}) \text{sinh} \rho(\varphi-\varphi_{0})-\text{sech}\rho( \varphi-\varphi_{0})\bigg], \label{max10}
\end{eqnarray}
\end{subequations}
being $C_{1}$ and $C_{2}$ arbitrary constants. Substituting the expresssions for $\phi_{11}$, $\phi_{12}$ and the explicit form of $\text{Re}F(\hat{A}_{0})$ in (\ref{max8}) we have 

\begin{eqnarray}
\phi_{1} &=& \dfrac{1}{\rho}\left[C_{2}+\dfrac{1}{2}\left(v\varphi_{0T}+\sigma_{0T}\right)\right]\text{sech}\rho (\varphi-\varphi_{0})\nonumber\\
&+& \left[C_{1}+\dfrac{3C_{2}}{2}(\varphi-\varphi_{0})+\dfrac{1}{2}(\varphi-\varphi_{0})(v\varphi_{0T}+\sigma_{0T})\right]\nonumber\\
&{\times}& \text{sech} \rho (\varphi-\varphi_{0})\text{tanh}\rho (\varphi-\varphi_{0})\nonumber\\
&+& \dfrac{C_{2}}{2\rho}\text{sinh}\rho (\varphi-\varphi_{0})\text{tanh}\rho (\varphi-\varphi_{0}) \label{max0}
\end{eqnarray} 
The last term in Eq. (\ref{max0}) makes the soliton unbounded and, hence, it can be removed by assuming the arbitrary constant $C_{2}=0$. By using the boundary condition $\phi_{1}|_{\varphi=\varphi_{0}}=c$ and $\phi_{1\varphi}|_{\varphi=\varphi_{0}}=0$, we obtain 
\begin{equation}
\dfrac{1}{\rho}(v\varphi_{0T}+\sigma_{0T})=-c \quad \text{and} \quad C_{1}=0 \label{lala1}
\end{equation}
and consequently, Eq. (\ref{max0}) turns out to be
\begin{equation}
\phi_{1}=c[1-(\varphi-\varphi_{0})\text{tanh}\rho (\varphi-\varphi_{0})]\text{sech}\rho (\varphi-\varphi_{0}). 
\end{equation}
A similar procedure can be done to find the general solution for Eq. (\ref{max4}). Henceforth, we have that $\psi_{1}$ has the following solution
\begin{eqnarray}
\psi_{1} &=& C_{3}\psi_{11}+C_{4}\psi_{12}-\psi_{11}\int_{-\infty}^{\infty}\psi_{12}\text{Im}F(A_{0})d\varphi\nonumber\\ 
&+& \psi_{12}\int_{-\infty}^{\infty}\psi_{11}\text{Im}F(A_{0})d\varphi,\label{o1}
\end{eqnarray}
where $C_{3}$ and $C_{4}$ are arbitrary constants, in addition, we have that $\psi_{11}$ and $\psi_{12}$ are the homogeneous part of (\ref{max4}) and their explicit forms are:
\begin{subequations}
\begin{equation}
\psi_{11}=\text{sech}\rho(\varphi-\varphi_{0}),\label{o2}
\end{equation}
\begin{equation}
\psi_{12}=\dfrac{1}{2\rho}\left[\rho(\varphi-\varphi_{0})\text{sech}\rho(\varphi-\varphi_{0})+\text{sinh}\rho(\varphi-\varphi_{0})\right]\label{o3}
\end{equation}
\end{subequations} 
Finally, the explicit form for $\psi$ can be given replacing Eqs. (\ref{o2}),  (\ref{o3}) and the explicit form for $\text{Im} F(\hat{A}_{0})$ in Eq. (\ref{o1}). Therefore, we have

\begin{eqnarray}
\psi_{1} &=& \Bigg(C_{3}+\dfrac{C_{4}}{2}(\varphi-\varphi_{0})-\dfrac{\rho}{2}\Bigg\{(\varphi-\varphi_{0})\Bigg[\dfrac{\rho_{T}}{2\rho}(\varphi-\varphi_{0})-\varphi_{T}\Bigg]\\\nonumber
&+& \varphi_{0T}\text{tanh}\rho(\varphi-\varphi_{0})\nonumber\\
&+& \varphi_{0T}(\varphi-\varphi_{0})\text{sech}^{2}\rho(\varphi-\varphi_{0})\Bigg\}\Bigg)\text{sech}(\varphi-\varphi_{0})\nonumber\\
&+&\dfrac{C_{4}}{2\rho}\text{sinh}\rho(\varphi-\varphi_{0}) \label{o4}
\end{eqnarray}

\begin{figure}[H]
\begin{subfigure}{.5\textwidth}
  \centering
  \includegraphics[width=.9\linewidth]{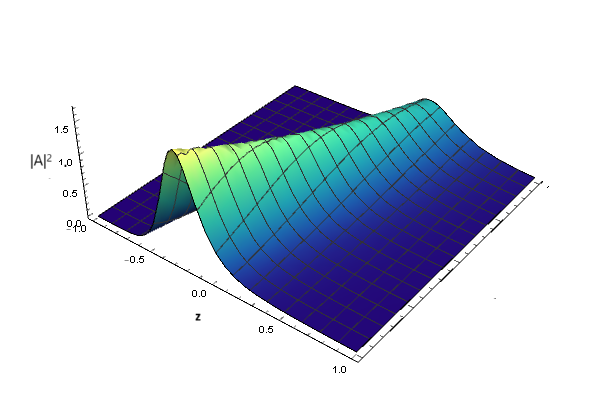}  
  \caption{}
  \label{sub1}
\end{subfigure}
\begin{subfigure}{.5\textwidth}
  \centering
  \includegraphics[width=.9\linewidth]{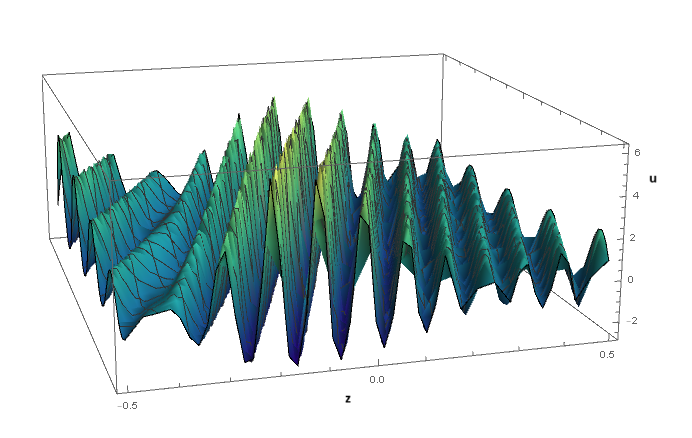}  
  \caption{}
  \label{sub2}
\end{subfigure}
\caption{(a) Envelope bright soliton at first order of $\gamma$. (b) Breathing soliton for the nerve density. }
\label{fig2}
\end{figure}

In the general solution for $\psi_{1}$, the term proportional to $\text{sinh} \rho(\varphi-\varphi_{0})$ can be removed by choosing $C_{4}=0$. We also get $C_{3}=0$ and $\varphi_{0T}=0$ upon using the boundary conditions $\psi_{1}|_{\varphi=\varphi_{0}}=0$ and $\psi_{1\varphi}|_{\varphi=\varphi_{0}}=0$. Consequently, resulting in the expression for $\sigma_{0T}$ provided by the explicit expression provided by Eq. (\ref{lala1}). On using the above results in Eq. (\ref{o4}), the final form of $\psi_{1}$ is obtained as 

\begin{equation}
\psi_{1}=\dfrac{\rho}{2}(\varphi-\varphi_{0})^{2}\text{sech}\rho(\varphi-\varphi_{0})
\end{equation}
Thus, the low amplitude excitation modelled by perturbed soliton solution due to axoplasmic fluid is given by 

\begin{align}
A &= \bigg[\rho \, \text{sech} \rho (\varphi-\varphi_{0})+c\gamma [1-(\varphi-\varphi_{0})\tanh\rho (\varphi-\varphi_{0})]\nonumber\\
&+ i\gamma\dfrac{\rho}{2}(\varphi-\varphi_{0})^{2}\text{sech} \rho (\varphi-\varphi_{0})\bigg]\exp[iv (\varphi-\varphi_{0})+i(\sigma-\sigma_{0})]  \label{p1}
\end{align}
and depicted in Fig. (\ref{sub2}), if we replace the last equation (\ref{p1}) in Eq. (\ref{wom4}) together with Eq. (\ref{wwmo0}) and Eq. (\ref{wwom1}) we will obtain the expression for the  nerve density: 

\begin{eqnarray}
u= 2(\chi \sin(\theta+\lambda)-\mu \cos(\theta +\lambda))\left[1+\frac{ap}{3k^2}(\chi \sin(\theta+\lambda)-\mu \cos(\theta +\lambda))\right] \label{upert}
\end{eqnarray}
with 
\begin{eqnarray}
\chi&=&\rho \text{sech}(\rho \Omega) +{\gamma}c(1-\Omega \text{tanh}(\rho \Omega)); \nonumber\\
\mu &=&{\gamma}\frac{\rho \Omega^2}{2} \text{sech}(\rho \Omega);\nonumber
\end{eqnarray}
and
\begin{equation}
\lambda = 2v\Omega + \sigma-\sigma_{0},\ \  \ \Omega= \varphi-\varphi_{0}.\ \ \  \nonumber
\end{equation}

The expression for the nerve density (\ref{upert}), depicted in Fig. \ref{sub2} represents the typical profile of a breathing soliton. This soliton like solution has been previously obtained by means of modulation instability analysis \cite{ra0, ra1}. Physiologically, this soliton solution has been associated with hyperpolarization and refractory periods. The adiabatic evolution of breathing soliton due to the elastic fluid could be translated in impulse blockage associated with mechanical anaesthesia. The expression corresponding to $|u|^{2}$ yields to the perturbed envelope bright soliton depicted in Fig. \ref{sub1}.

\subsection{Adiabatic evolution of dark soliton in liquid state of lipid bilayer}

As it was seen from the previous section, the propagation of bright solitons under perturbation is described by the adiabatic evolution of the soliton parameters; i.e. the soliton's height, velocity, position shift and phase shift. However, the non-vanishing boundary of dark solitons introduces serious complications once perturbative methods developed for bright solitons are applied \cite{was1, was2, was3, was4}. In this direction M. A. Ablowitz and collaborators proposed a method to study perturbed adiabatic evolution of dark solitons including shelves structures that arises around solitons \cite{o4}. We must point it out that in the present work we are going to restrict ourselves to the core soliton evolution, nevertheless we can extend the present study to the general case. Similar to the study of bright solitons in nerve channels we consider the dimensionless nonlinear Schrödinger equation with a damping term choosing the negative value of the dispersion term, that is

\begin{equation}
i\dfrac{\partial A}{\partial \tilde{t}}-\dfrac{1}{2}\dfrac{\partial^{2} A}{\partial z^{2}}+|A|^{2}A+i\gamma A=0 \label{lap1},
\end{equation}

A non-vanishing boundary value at infinity; i.e., $|A|$ need not to be zero as $z$ tends to large values. The effect that perturbation has on the behaviour of the solution at infinity is independent of any local phenomena such as pulses that do not decay at infinity; i.e., dark solitons. In the case of a continuous wave background, which is relevant to perturbation problems with dark solitons we have that $A_{zz} \to 0$ as $z\to \pm\infty$, and the evolution of the background at either end $A \to A^ {\pm}(\tilde{t})$ is given by the equation

\begin{equation}
i\dfrac{d}{d\tilde{t}}A^{\pm}+|A^ {\pm}|^ {2}A^{\pm}=-i\gamma A ^{\pm}.\label{lap2}
\end{equation}
We write $A^ {\pm}(\tilde{t})=\alpha^{\pm}(\tilde{t})e^{i\phi^{\pm}(\tilde{t})}$, where $\alpha^{\pm}(\tilde{t})>0$ and $\phi^{\pm}(\tilde{t})>0$ and $\phi^{\pm}(t)$ are both real functions of $\tilde{t}$. Then, the real and imaginary parts of the above equations are given by

\begin{subequations}
\begin{equation}
\dfrac{d\phi^{\pm}(\tilde{t})}{d\tilde{t}}=(\alpha^{\pm}(\tilde{t}))^{2}, \label{lap3}
\end{equation}
\begin{equation}
\dfrac{d\alpha^{\pm}(\tilde{t})}{d\tilde{t}}=\gamma {\alpha^{\pm}(\tilde{t})},\label{lap4}
\end{equation}
\end{subequations}
which are equations that completely describe the adiabatic evolution of the background under the influence of the perturbation. In addition, we consider that at $\tilde{t}=0$ $\alpha^{+}(0)=\alpha^ {-}(0)$, then, since $\alpha^{\pm}(\tilde{t})$ satisfy the same equation, the evolution is the same for all $t$. Hence, $\alpha^{+}(\tilde{t})=\alpha^ {-}(\tilde{t})\equiv \alpha_{\infty}(\tilde{t})$. In addition considering the phase difference $\Delta \phi_{\infty}(t)=\phi^{+}(\tilde{t})-\phi^{-}(\tilde{t})$ which is related to the depth of dark soliton; here $\phi^ {\pm}(\tilde{t})$ represents the phase as $z\pm \infty$, respectively, 
\begin{equation}
\dfrac{d}{d\tilde{t}}\alpha_{\infty}=-\gamma \alpha_{\infty} \quad \text{and} \quad \dfrac{d}{d\tilde{t}}\Delta \phi_{\infty}=0.\label{lap5}
\end{equation} 
Thus, we conclude that the magnitude of the background evolves adiabatically, nevertheless the phase difference remains unaffected by the perturbation. \\
Let us first consider the fast evolution of the background phase as

\begin{equation}
A=\alpha e^{-\int_{0}^{\tilde{t}}\alpha_{\infty}(s)^{2}ds},\label{lap6}
\end{equation}
so, Eq. (\ref{lap7}) becomes
\begin{equation}
i\alpha_{\tilde{t}}-\dfrac{1}{2}\alpha_{zz}+\left(|\alpha|^ {2}-\alpha^{2}_{\infty}\right)\alpha=-i\gamma \alpha.\label{lap7}
\end{equation}
The dark soliton solution of the unperturbed equation is given by
\begin{equation}
\alpha_{s}(z, \tilde{t})=\left(A+iB\tanh[B(z-A\tilde{t}-z_{0})]\right)e^{i\sigma_{0}},\label{lap8}
\end{equation}
where the core parameters of the soliton, $A$, $B$, $z_{0}$ and $\sigma_{0}$ are all real, the magnitude of the background is $(A^ {2}+B^ {2})^{1/2}=\alpha_{\infty}$ and the phase difference across the soliton is $2\tan^{-1}(B/A)$, $A\neq 0$. It is worth to point out that when $A=0$ we have the expression of a black soliton of phase difference of $\pi$.
We write the solution in terms of the amplitude and phase $\alpha=qe^{i\phi}$, where $q$ and $\phi$ are both real functions of $\tilde{t}$ and $z$ and Eq. (\ref{lap7}) becomes
\begin{equation}
iq_{\tilde{t}}-\phi_{\tilde{t}}q-\dfrac{1}{2}\left(q_{zz}+2i\phi_{z}q_{z}+q(i\phi_{zz}-\phi^{2}_{z})\right)+\left(|q|^ {2}-\alpha_{\infty}^{2}\right)q=-i\gamma q, \label{q1}
\end{equation}
again by implementing a multi-scale analysis we introduce a slow scale variable $T=\gamma{\tilde{t}}$ with the parameters $A$, $B$, $z_{0}$ and $\sigma_{0}$ being the functions of $T$ and expand $q$ and $\phi$ as series of $\gamma$, that is, $q(z, T, \tilde{t})=q_{0}+\gamma q_{1}+O(\gamma^ {2})$ and $\phi(z, T, \tilde{t})=\phi_{0}+\gamma\phi_{1}+O(\gamma^{2})$ at first order we have
\begin{subequations}
\begin{equation}
q_{0\tilde{t}}=\dfrac{1}{2}\left(2\phi_{0z}q_{0z}+q_{0}\phi_{0zz}\right),\label{q2}
\end{equation}
\begin{equation}
\phi_{0\tilde{t}}q_{0}=-\dfrac{1}{2}\left(q_{0zz}-\phi_{0z}^{2}q_{0}\right)+\left(|q_{0}|^{2}-\alpha_{\infty}^{2}\right)q_{0},\label{qq2}
\end{equation}
\end{subequations}
with the general dark soliton unperturbed solution 
\begin{subequations}
\begin{equation}
q_{0}=\left(A(T)+B(T)^{2}\tanh^ {2}(\xi)\right)^ {1/2},\label{qqq2}
\end{equation}
\begin{equation}
\phi_{0}=\tan^{-1}\left[\dfrac{B(T)}{A(T)}\tanh(\xi)\right]+\sigma_{0}(T),\label{q3}
\end{equation}
\end{subequations}
where $\xi=B\left(z-\int^{z}_{0}A(\epsilon s)ds-z_{0}(T)\right)$. For a dark soliton, the expressions (\ref{qqq2}) and (\ref{q3}) are taken to be 
\begin{subequations}
\begin{equation}
q_{0}(z, T, \tilde{t})=\alpha_{\infty} \tanh [\alpha_{\infty}(z-z_{0}(T))],\label{q4}
\end{equation}
\begin{equation}
\phi_{0}(z, T, \tilde{t})=\sigma_{0}(T).\label{q5}
\end{equation}
\end{subequations}
For $O(\gamma)$, we have
\begin{subequations}
\begin{equation}
q_{1\tilde{t}}=\dfrac{1}{2}\left[2(\phi_{0z}q_{1z}+q_{0z}\phi_{1z})+q_{0}\phi_{1zz}+q_{1}\phi_{0zz}\right]-q_{0}- q_{0T}\label{q6}
\end{equation}
\begin{equation}
\phi_{1\tilde{t}}q_{0}=-q_{1}\phi_{0\tilde{t}}-\dfrac{1}{2}\left[q_{1zz}-(2\phi_{0z}\phi_{1z})q_{0}-\phi_{0z}^ {2}q_{1}\right]+3q_{0}^{2}q_{1}-\alpha_{\infty}^ {2}q_{1}-\phi_{0T}q_{0}.\label{q7}
\end{equation}
\end{subequations}
Let us consider that the black pulse satisfies boundary conditions, Eq. (\ref{lap5}). Thus, the slow evolution terms are
\begin{equation}
q_{0T}=-z_{0T}q_{0z}\quad \text{and} \quad \phi_{0T}=\sigma_{0T}.\label{q8}
\end{equation}
If we look for stationary solution, $q_{1\tilde{t}}=\phi_{1\tilde{t}}=0$, and note that $\phi_{0z}=\phi_{0zz}=0$, then Eqs. (\ref{q6}) and (\ref{q7}) reduce to

\begin{subequations}
\begin{equation}
-\dfrac{1}{2}q_{1zz}+3q_{0}^{2}q_{1}-\alpha_{\infty}^ {2}q_{1}-\phi_{0T}q_{0}=0,\label{s1}
\end{equation}
\begin{equation}
\dfrac{1}{2}q_{0}\phi_{1zz}+q_{0z}\phi_{1z}-q_{0}+z_{0T}q_{0z}=0. \label{q9}
\end{equation}
\end{subequations}
Thus, the expression for a black soliton is given by 
\begin{equation}
q_{1}=\dfrac{\sigma_{0T}}{4\alpha_{\infty}}[\sinh(2\alpha_{\infty}(z-z_{0}))+2\alpha_{\infty}(z-z_{0})]\text{sech}^{2}(\alpha_{\infty}(z-z_{0}))\label{s2}
\end{equation}
whose asymptotic behaviour 
\begin{equation}
q_{1}^{\pm}=\pm \dfrac{\sigma_{0T}}{2\alpha_{\infty}}, \label{s3}
\end{equation}
we kept the superscript $\pm$ to indicate the asymptotic behaviour, i.e., $z\to \pm \infty$, correspondingly. Similarly, we can obtained the explicit expression for $\phi_{1}$ 
\begin{eqnarray}
\phi_{1}&=&\dfrac{1}{\alpha_{\infty}^{2}}\left[\dfrac{\alpha_{\infty}^{2}(z-z_{0})^{2}}{2}-\alpha_{\infty}(z-z_{0})\text{coth}(\alpha_{\infty}(z-z_{0})\right]+\dfrac{C_{1}}{\alpha_{\infty} ^{3}}\left[\alpha_{\infty}(z-z_{0})-\text{coth}(\alpha_{\infty}(z-z_{0}))\right]\nonumber\\
&-& z_{0T}(z-z_{0})+C_{2},\label{mad0}
\end{eqnarray}
similar to the case of $q_{1}$ provided by Eq. (\ref{s2}) we may choose $C_{1}$ and $C_{2}$ as free parameters to remove exponential growth and to maintain the antisymmetric property of $\phi$.  
To find the slowly evolving parameters $\sigma_{0}(T)$ and $z_{0}(T)$ we derive the equations for growth of the shelf from the perturbed conservation laws associated with the first order perturbation of NLSE and the asymptotic behaviour of its solutions. The shelf is associated with the asymptotic parameters $q_{1}^{\pm}$ and $\phi_{1z}^{\pm}$, which are expressed in terms of $\sigma_{0T}$ and $z_{0T}$. We use the Hamiltonian $H$, the energy $E$, the momentum $P$ and the centre of energy $R$ given for the following expressions
\begin{subequations}
\begin{equation}
H=\int_{-\infty}^{\infty} \left[\dfrac{1}{2}\bigg|\dfrac{\partial \alpha}{\partial z}\bigg|^{2}+\dfrac{1}{2}(\alpha_{\infty}^{2}-|\alpha|^{2})^{2}\right]dz;
\end{equation}
\begin{equation}
E=\int_{-\infty}^{\infty}[\alpha_{\infty} ^{2}-|\alpha|^{2}]dz;
\end{equation}
\begin{equation}
P=\int_{-\infty}^{\infty} \text{Im} [\alpha \alpha^{*}_{z}]dz;
\end{equation}
\begin{equation}
R=\int_{-\infty}^{\infty} z(\alpha_{\infty}^{2}-|\alpha|^{2})dz
\end{equation}
\end{subequations}
For the unperturbed NLSE the first three integrals are conserved quantities while the last can be written in terms of the momentum. The evolution of the last equations can be easily obtained turning to be 
\begin{subequations}
\begin{equation}
\dfrac{dH}{d\tilde{t}}=\gamma \left(E\dfrac{d}{dT}\alpha_{\infty}^{2}+2\text{Re}\int_{-\infty}^{\infty}F[\alpha]\alpha_{\tilde{t}}^{*}dz\right),
\end{equation} 
\begin{equation}
\dfrac{dE}{d\tilde{t}}=2\gamma \text{Im}\int_{-\infty}^{\infty}\left(F[\alpha_{\infty}]\alpha_{\infty}-F[\alpha]\alpha ^{*}\right)dz,
\end{equation}
\begin{equation}
\dfrac{dP}{d\tilde{t}}=2\gamma\text{Re}\int_{-\infty}^{\infty}F[\alpha]\alpha_{\tilde{t}}^{*}dz,
\end{equation}
\begin{equation}
\dfrac{dR}{d\tilde{t}}=-P+2\gamma \text{Im}\int_{-\infty}^{\infty}z\left(F[\alpha_{\infty}]\alpha_{\infty}-F[\alpha]\alpha ^{*}\right)dz.
\end{equation}
\end{subequations}
Thus, let us begin with the perturbed conservation of energy
\begin{equation}
\dfrac{d}{d\tilde{t}}\int_{-\infty} ^{\infty} [\alpha_{\infty}^{2}-|\alpha|^{2}]dz=2\gamma \text{Im} \int_{-\infty} ^{\infty} (F[\alpha_{\infty}]\alpha_{\infty}-F[\alpha]\alpha^{*})dz.
\end{equation}
For our case we have $\alpha=qe ^{i\phi}$, $F[u]=i\alpha$, $Z=z-z_{0}$ expanding $q=q_{0}+\gamma q_{1}+...$ and taking terms of first order of $\gamma$, we have
\begin{equation}
\dfrac{d}{d\tilde{t}}\int_{-\alpha_\infty \tilde{t}} ^{\alpha_\infty \tilde{t}} q_{0}q_{1} dZ=-\int_{-\infty} ^{\infty} q_{0} ^{2}dZ. 
\end{equation}
This is an equation for the change in energy caused by the propagation of the shelf. The integration will be over $T \in [-\alpha_\infty z, \alpha_\infty z]$. Thus, we're integrating over the inner region around the soliton. Since $q_{0}$ and $q_{1}$ are only functions of $Z$ and $T$, we can apply the fundamental theorem of calculus to arrive at
\begin{equation}
\alpha_{\infty}[q_{1}(\alpha_{\infty})q_{0}(\alpha_{\infty})+q_{1}(-\alpha_{\infty})q_{0}(-\alpha_{\infty})]=2\alpha_{\infty}
\end{equation}
And, for large $z$ we take $q_{0}\to \pm \alpha_{\infty}$ and $q_{1}\to q_{1} ^{\pm}$ and the asymptotic values for $q_{1}^{\pm}$, thus we arrive to the expression for $\sigma_{0T}$
\begin{equation}
\sigma_{0T}=2\alpha_{\infty}
\end{equation}
Now, considering the modified conservation of momentum
\begin{equation}
\dfrac{d}{d\tilde{t}}\text{Im}\int_{-\infty} ^{\infty} \alpha \alpha_{z}^{*}dz=2\gamma\text{Re} \int_{-\infty}^{\infty}F[\alpha]\alpha_{z} ^{*}dz.
\end{equation}
Again, considering $u=qe^{i\phi}$, $F[\alpha]=-iq$, $Z=z-z_{0}$ and use the perturbation expansion for $\alpha$ up the first order of $\gamma$ we have
\begin{equation}
-\dfrac{d}{d\tilde{t}}\int_{-\infty}^{\infty} [\phi_{0Z}q_{0}^{2}+\gamma (2\phi_{0Z}q_{0}q_{1}+\phi_{1z}q_{0} ^{2})]dZ=2\gamma \text{Re} \int_{-\infty}^{\infty} i q_{0}q_{0Z}dZ, 
\end{equation}
which, using $\phi_{0Z}=0$, reduces in the same way as the conservation of energy to 
\begin{equation}
\phi_{1z} ^{+}+\phi_{1z} ^{-}=0
\end{equation}
By substituting the asymptotic approximations (\ref{s3}) found earlier for $\phi_{1z}^{\pm}$, we arrive at the expression $z_{0}$
\begin{equation}
z_{0T}=0
\end{equation}
Finally, we obtain the expression for the first order perturbed dark soliton $A$ is given by 
\begin{align}
A &= \left(\alpha_{\infty}\tanh[\alpha_{\infty}(z-z_{0})]+\gamma \dfrac{\sigma_{0T}}{4\alpha_{\infty}}[\sinh(2\alpha_{\infty}(z-z_{0}))+2\alpha_{\infty}(z-z_{0})]\text{sech}^{2}(\alpha_{\infty}(z-z_{0}))\right)\times\nonumber\\
& \exp\bigg(i\sigma_{0}+i\gamma \dfrac{1}{\alpha_{\infty}^{2}}\left[\dfrac{\alpha_{\infty}^{2}(z-z_{0})^{2}}{2}-i\alpha_{\infty}(z-z_{0})\text{coth}(\alpha_{\infty}(z-z_{0}))\right]\nonumber\\ 
& +i\gamma \dfrac{C_{1}}{\alpha_{\infty} ^{3}}\left[\alpha_{\infty}(z-z_{0})-\text{coth}(\alpha_{\infty}(z-z_{0}))\right]-i\gamma z_{0T}(z-z_{0})+i\gamma C_{2}\bigg)\exp\left(\int_{0} ^{\tilde{t}}\alpha_{\infty}(s) ^{2}ds\right). \label{joc0}
\end{align}

\begin{figure}[ht]
\centering
\includegraphics[width=9cm]{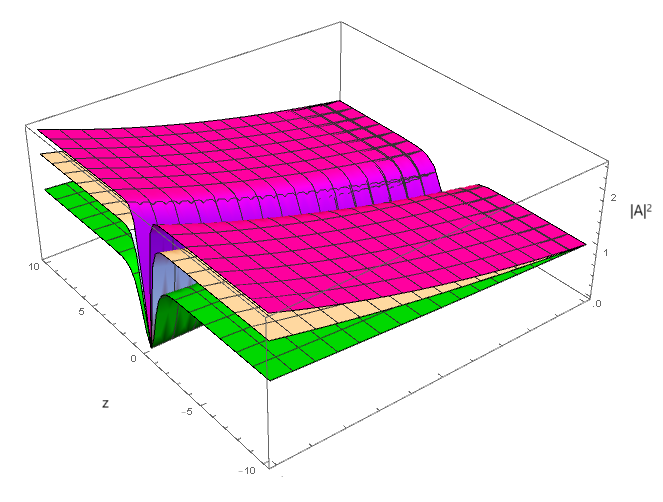}
\caption{Representation of first order perturbed dark soliton for different values of $\gamma$.} \label{ig4}
\end{figure}

In order to obtain the expression for the density $u$ we substitute (\ref{joc0}) in Eq. (\ref{wom4}) together with Eq. (\ref{wwmo0}) and Eq. (\ref{wwom1}) to obtain the expression:
\begin{equation}
u= 2 M R \cos(H)+ a\left[\frac{p M^2 R^2}{3k^2}\right]\left(e^{2H}-\cos(2H)\right)\label{mag11}
\end{equation}
with
\begin{eqnarray}
M&=&\alpha_{\infty}\tanh(\Theta)+\frac{\gamma \sigma_{0T}}{4\alpha_{\infty}}\left(\sinh(2\Theta)+2\Theta\right)\sinh^2\Theta; \nonumber\\
H&=&\frac{\gamma}{\alpha_{\infty}^2} \Theta \coth(\Theta); \nonumber\\
R&=& \exp\left(-\int_0^{\bar{t}}\alpha_{\infty}(s)^2 ds\right);\nonumber\\
G&=&\sigma_{0}+\frac{\gamma \Theta^2}{2\alpha_{\infty}^2} + 
\frac{\gamma C_{1}}{\alpha_{\infty}^3}(\Theta -\coth(\Theta))-\frac{\gamma \Theta z_{0T}}{\alpha_{\infty}} + \gamma C_{2}\nonumber\\
\Theta &=& \alpha_{\infty}(z-z_{0}).
\end{eqnarray}
Again, as it was expected for the case of $|u|^{2}$ provided by Eq. (\ref{mag11}) we obtain the profile of the perturbed envelope dark soliton depicted in Fig. \ref{ig4}. Similar to the case of the envelope bright soliton for $|u| ^{2}$ provided by Eq. (\ref{upert}) for lipid bilayers.  One of the many advantages of the method developed for the liquid state of lipid bilayers is the possibility of the introduction of the asymmetrical properties of the cytolemma, that is the inner and outer leaflets, one being more fluid than the other. As it is known the main reason for such an asymmetry is the fact that cytolemma separates domains of different biochemical composition and different physical properties, namely, cytoplasm and extracellular fluid. In addition, there have been shown that  ionic transmembrane shifts must be accompanied by water movement across the membrane, which contributes to the observed transient cellular expansions and might also play a functional role in neuronal activation via complex interactions with the cytoskeleton \cite{lebi}. The latter can be modelled as the shelves that forms around the perturbed dark soliton.  An extended analysis about the adiabatic evolution of dark solitons, together with shelves and its implications in lipid membranes will be carried out in a future work. 

\section{Conclusion}
Both membrane and lipid bilayers, while closely assembled, exhibit different levels of rigidity, which significantly influence the nonlinear effects in their mechanical distortions. These mechanical characteristics can be effectively studied using direct approaches that introduce them as generalized perturbation functions, thus we can extend the present study to another mechanical perturbations. Another distinguishing aspect is the exchange of matter between the cell's interior and its environment. The addition or reduction of specific molecules inevitably alters the structural, mechanical, and electrical properties of the membrane. This exchange leads to a phase transition where the lipid bilayer exists not only in ordered gel or disordered liquid states but also in a distinct third phase known as the ordered liquid phase, which coexists within a relatively wide temperature range around its melting point.
The structures of ordered liquid phases above and below the melting temperature are similar, translating in greater rigidity to the membranes compared to the disordered liquid phase, while still allowing for molecular migration across the membrane, a crucial feature of ordered gel systems. Multiple phases can coexist simultaneously within a typical biomembrane under physiological conditions. This exchange process can be analyzed within the framework of the perturbed nonlinear Schrödinger equation (NLSE) or by drawing an analogy to phase transitions in systems like Bose-Einstein condensates. The perturbed NLSE also enables the study of asymmetrical properties within the cytolemma, particularly between its inner and outer leaflets, where one leaflet may exhibit greater fluidity. This asymmetry arises because the cytolemma separates domains with different biochemical compositions and physical properties, such as the cytoplasm and extracellular fluid.
Moreover, the model facilitates the investigation of adiabatic evolution and symmetry properties of orthodromic and antidromic impulses in nerves. Many features of nerves and other biomembranes remain elusive, but nonlinear, low-amplitude excitations provide insights into these complex systems.

\end{document}